\documentclass{aa}
\input{psfig.sty}

\def\e20{$\times 10^{20}$}
\def\emin12{$\times 10^{-12}$}

\def\>{$>$}
\def\<{$<$}

\def\newline{\hfil\break}
\def\mincir{\ \raise -2.truept\hbox{\rlap{\hbox{$\sim$}}\raise5.truept  
\hbox{$<$}\ }}                        
\def\magcir{\ \raise -2.truept\hbox{\rlap{\hbox{$\sim$}}\raise5.truept  
\hbox{$>$}\ }}                        

\def\ecs{erg cm$^{-2}$ s$^{-1}$}
\def\es{erg s$^{-1}$}
\def\cm2{cm$^{-2}$}
\def\sp{$\,$}

\catcode`\@=11
\def\gsim{\ifmmode{\mathrel{\mathpalette\@versim>}}
    \else{$\mathrel{\mathpalette\@versim>}$}\fi}
\def\lsim{\ifmmode{\mathrel{\mathpalette\@versim<}}
    \else{$\mathrel{\mathpalette\@versim<}$}\fi}
\def\@versim#1#2{\lower 2.9truept \vbox{\baselineskip 0pt \lineskip 
    0.5truept \ialign{$\m@th#1\hfil##\hfil$\crcr#2\crcr\sim\crcr}}}
\catcode`\@=12

\begin{document}

\title{Nuclear and global X-ray properties of LINER galaxies:
$Chandra$ and $BeppoSAX$ results for Sombrero and NGC 4736}

\author{S. Pellegrini \inst{1}, G. Fabbiano \inst{2}, F. Fiore \inst{3},
G. Trinchieri \inst{4}, A. Antonelli \inst{3}}

\offprints{S. Pellegrini}

\institute{
Universit\`a di Bologna, Dipartimento di Astronomia, via Ranzani 1, I-40127
Bologna, Italy\\
\email{pellegrini@bo.astro.it}
\and
Harvard-Smithsonian Center for Astrophysics, 60 Garden Street, Cambridge, 
MA 02138, USA\\
\email{pepi@head-cfa.harvard.edu}
\and
Osservatorio Astronomico di Roma, via Frascati 33, I-00044 Monteporzio 
Catone, Italy\\
\email{fiore@quasar.mporzio.astro.it,angelo@coma.mporzio.astro.it}
\and
Osservatorio Astronomico di Brera, via Brera 28, I-20121 Milan, Italy\\
\email{ginevra@brera.mi.astro.it}
}

\date{Received July 23, 2001; accepted October 22, 2001}

\abstract{ We report on the 0.1--100 keV $BeppoSAX$ observations of
two nearby LINER galaxies, Sombrero and NGC 4736.  $Chandra$ ACIS-S
observations supplement this broad-beam spectral study with a high
resolution look into the nuclear region, and show a dominating central
point source in Sombrero and a complex X-ray binary
dominated/starburst region in NGC 4736.  A compact non-thermal radio
source, present in the nucleus of both galaxies, coincides with the
central source in Sombrero, while in NGC 4736 its X-ray counterpart is
a much fainter point source, not the brightest of the central
region. On the basis of these and other results, we conclude that the
LINER activity is linked to the presence of a low luminosity AGN in
Sombrero and to a recent starburst in NGC 4736, and that $Chandra$'s
spectroscopic capabilities coupled to high resolution imaging are
essential to establish the origin of the nuclear activity.  
}

\authorrunning{S. Pellegrini et al.}

\titlerunning{{\it Chandra} and {\it Beppo}SAX views of Sombrero
and NGC 4736}

\maketitle

\section{Introduction}

Optical spectroscopic surveys showed that low ionization nuclear
emission line regions (LINERs, Heckman 1980) are very common among
ellipticals and early type spirals (Ho et al. 1997). It is also
currently believed that most galactic spheroids host supermassive
black holes, based on {\it Hubble Space Telescope} ($HST$) stellar or
ionized gas spectroscopy coupled to dynamical modeling (e.g.,
Richstone et al. 1998).  So, it has become natural to ask whether AGN
activity stops at Seyfert galaxies or extends down to lower
luminosities, including LINERs.  The answer to this question has
important consequences for the history of accretion in the Universe
(e.g., Haiman \& Menou 2000). The LINER emission, though, need
not be powered by AGN-like activity: collisional ionization by
shocks and/or photoionization by the UV radiation from hot, young
stars were also suggested as possible mechanisms (Terlevich et
al. 1992, Alonso-Herrero et al. 2000). Observational support for this
interpretation, in some cases, comes from optical and ultraviolet
$HST$ observations (e.g., Maoz et al. 1995) and from infrared studies
(Larkin et al. 1998).  

An important way of discriminating between different emission
mechanisms is to observe LINERs in hard X-rays, where the AGN presence
is easily identifiable, even when it is obscured by intervening matter
(e.g., the LINER NGC 6240 turned out to be an obscured, high
luminosity AGN from hard X-ray observations; Vignati et al. 1999).
For this reason we observed Sombrero (NGC 4594) and NGC 4736 with
$BeppoSAX$ over 0.1--100 keV, as they are early type spirals hosting
two of the closest examples of LINER activity (Table 1). Thanks to
their proximity, they are excellent targets to explore the
questions raised above and have been subject to thorough studies at various
wavelengths, including high angular resolution observations with
$HST$, the $ROSAT$ HRI and the VLA.

However, a $BeppoSAX$ study is severely limited by the available
angular resolution, as are all the reports so far on the hard X-ray
properties of LINERs based on spectra from a few arcminute
beams. These show power laws that could equally come from an active
nucleus or the collective emission of X-ray binaries [see the reviews
by Fabbiano (1996) and Serlemitsos, Ptak \& Yakoob (1996); and Ptak et
al. (1999)].  When the analysis of the $BeppoSAX$ data was complete, the
$Chandra$ observations of our target galaxies became available in the
public archive, providing us with the unique opportunity to
supplement our study with a very high resolution look into the
nuclear regions.  We therefore added the analysis of these data to our
work.  The $Chandra$ observation of Sombrero was reported as part of a
snapshot survey of faint nearby AGNs (Ho et al. 2001); we considered
more in detail those data here. After submission of this paper, a
preprint appeared on the astro-ph archive discussing in detail the
overall $Chandra$ observation of NGC 4736 (Eracleous et al. 2001);
our analysis is limited to the centermost region only.

\subsection{Properties of the target galaxies}

At the nucleus of Sombrero, identified as a LINER (Heckman 1980), high
resolution $HST$ spectroscopy confirmed the presence of a central dark
object of $10^9 M_{\odot}$ and also revealed a faint broad $H\alpha$
component with full width at zero intensity of $\approx 5200$ km
s$^{-1}$ (Kormendy et al. 1996). This nucleus hosts a compact and
variable radio continuum source (Thean et al. 2000) and is a pointlike
source in an $HST$ image at 3400\AA\sp (Crane et al. 1993).

In the X-ray, a $ROSAT$ HRI image showed a pointlike central source,
clumpy emission associated with the disk and diffuse or not resolved
emission from the bulge (Fabbiano \& Juda 1997, hereafter FJ).  How
much of this diffuse emission comes from a hot medium and how much
from the integrated emission of evolved stellar sources could not be
established.  $ROSAT$ PSPC spectra of the central circles of
$1^{\prime}$ and $2^{\prime}$ radius indicate a steeper photon index
(from $\Gamma=1.76 ^{+0.44}_{-0.36}$ to $\Gamma=1.84
^{+0.41}_{-0.29}$) for the larger aperture, which FJ interpreted as an
indication that the extended HRI emission is relatively soft.
Nicholson et al. (1998)	 obtain $\Gamma =1.63\pm 0.05$ and a column
density $N_H=5.3\times 10^{20}$ \cm2\sp from the combined $ASCA$ and
$ROSAT$ spectra; allowing for contamination by soft emission, the
column density to the power law is $2.9\times 10^{21}$ \cm2.  Ptak et
al. (1999), from combined $ASCA$ and $ROSAT$ data, find thermal
emission with very low abundance ($<0.1 Z_{\odot}$) coupled to a power
law of $\Gamma= 1.97^{+0.33}_{-0.29}$ absorbed by an excess
$N_H<1.7\times 10^{22}$\cm2\sp over the Galactic value.  Recently, Ho
et al. (2001) report the first results from an arcsecond resolution
snapshot survey of nearby galaxies performed by $Chandra$.  In their
morphological classification into four classes of the X-ray images
obtained, Sombrero belongs to the class of objects with a ``dominant
nuclear source''.

NGC 4736, a LINER2 galaxy (Ho et al. 1997), shows various pieces of
evidence suggesting star formation that took place in its recent
past. Balmer absorption lines and strong CO absorption indicate a
nuclear starburst that occurred $\sim 1$ Gyr ago (Wong \& Blitz
2000). A bright optical ring is prominent in HI (Mulder \& van Driel
1993), molecular gas (Gerin, Casoli \& Combes 1991) and H$\alpha$ (Pogge 1989).
A younger stellar population is also revealed by 6cm radio continuum
observations of compact HII regions associated with hot, young stars
and young supernova remnants (Turner \& Ho 1994) and by near-infrared
spectra (Larkin et al. 1998). The nuclear region hosts a compact
non-thermal radio continuum source (Turner \& Ho 1994) and a bright UV
point source revealed by $HST$ (Maoz et al. 1995).

A $ROSAT$ PSPC pointing showed within the central few kpc a compact
source plus an extended distribution of hot gas ($kT\sim 0.3$ keV)
contributing 30--35\% of the total observed 0.1--2 keV flux (Cui,
Feldkhun \& Braun 1997).  $ROSAT$ HRI data showed 12 discrete sources
superimposed on the optical disk and ring regions, likely X-ray
binaries (hereafter XRBs), supernova remnants or recent supernovae;
the brightest source was by far the galactic ``nucleus'' (Roberts,
Warwick \& Ohashi 1999, hereafter R99).  The combined PSPC and $ASCA$
spectrum (R99) is composed of a hard continuum, equally well modeled
with a power law (with $\Gamma\sim 1.7$) or bremsstrahlung emission
(with $kT\sim 8$ keV).  The soft thermal emission was modeled with a
two-temperature plasma (of $kT=0.1$ and 0.6 keV) and solar abundance.
A Fe K$\alpha$ line may also be present at 6.81$^{+0.13}_{-0.28}$ keV.

\begin{table*}
\caption[] { General galaxy properties}
\begin{flushleft}
\begin{tabular}{ l  rl  l  l l  l  l l  l l  l }
\noalign{\smallskip}
\hline
\noalign{\smallskip}
Galaxy & Type$^a$ & RA$^a$  & Dec$^a$ & d$\, ^b$ &$B_{\rm T}^0 \, ^a$ & 
log$L_{\rm B} \, ^c$ & $N_{\rm H,Gal} \, ^d$ \\
      &  &(J2000) &(J2000) &  (Mpc)     &   (mag)           & ($L_{\odot}$)  &  (cm$^{-2}$)        \\
\noalign{\smallskip}
\hline
\noalign{\smallskip}
Sombrero & Sa & $12^h 39^m 59^s \hskip -0.1truecm .4$ & $-11^{\circ} 37^{\prime} 23^{\prime\prime} $ & 9.4 & 8.38 & 10.79  &  3.5$\times 10^{20}$ \\
NGC 4736 & Sab & $12^h 50^m 53^s \hskip -0.1truecm .6$ & $+41^{\circ} 07^{\prime} 10^{\prime\prime} $ & 5.9 & 8.75 & 10.23  &  1.4$\times 10^{20}$ \\
\noalign{\smallskip}
\hline
\end{tabular} 
\end{flushleft}
\bigskip
$^a$ from de Vaucouleurs et al. (1991).  $B_{\rm T}^0$ is the total B 
magnitude, corrected for extinction. 

$^b$ distance from Ajhar et al. (1997) for Sombrero and 
Sch\"oniger \& Sofue (1994) for NGC 4736.

$^c$ total B-band luminosity, derived using the indicated distance and 
$B_{\rm T}^0$. 

$^d$ Galactic neutral hydrogen column density from Stark et al. (1992).

\end{table*}

\section {Spatial analysis}

\subsection{$BeppoSAX$ observations }

Sombrero and NGC 4736 were observed with 
 the Low Energy Concentrator Spectrometer (LECS, Parmar et
al. 1997), the Medium Energy Concentrator Spectrometer (MECS, Boella
et al. 1997) and the Phoswich Detector system (PDS).
The latter is a collimated instrument, operating in rocking mode, that
covers the 13--300 keV energy band. It has a triangular response with
FWHM of $\sim 1^{\circ}\hskip -0.1truecm .3 $ (Frontera et al. 1997).
Its data were reduced using the variable risetime
threshold technique to reject particle background (Fiore, Guainazzi
\& Grandi 1999).
The journal of these observations is given in Table 2.  

\begin{table*}
\caption[]{$BeppoSAX$ observation Log}
\begin{flushleft}
\begin{tabular}{lllllllllllll}
\noalign{\smallskip}
\hline
\noalign{\smallskip}
Galaxy & Date        & \multicolumn{3}{l}{Exposure time$^a$ (ks)}  &  & 
\multicolumn{3}{l}{Count Rate$^b$ ($ 10^{-2}$ ct/s)}                 \\
   \cline{3-5}\cline{7-9}
   &     &  LECS   & MECS   & PDS       &         & LECS & MECS & PDS \\
   &    &       &      &     &  & 0.1--4 keV & 1.7--10 keV & 13-30 keV \\
\noalign{\smallskip}
\hline
\noalign{\smallskip}
Sombrero & 2000 Jun 29  & 23.635 & 72.087 & 32.759 & & 1.66$\pm 
0.11$  & 2.82 $\pm 0.08$ & $4.32\pm 1.35$     \\
NGC 4736 & 2000 Dec 29  & 24.356 & 76.182 & 34.677 & & $1.55\pm 
0.09$ & $2.29\pm 0.06$ &    \\ 
\noalign{\smallskip}
\hline
\end{tabular} 
\end{flushleft}
\bigskip

$^a$ On-source net exposure time. The LECS exposure time is
considerably shorter than the MECS one, because the LECS
can operate only when the spacecraft is not illuminated by the Sun.

$^b$ Background subtracted count rates, with photon counting
statistics errors, from extraction radii of $6^{\prime}$ for
Sombrero [with the bright G0 star (Sect. 2.1) 
removed] and $4^{\prime}$ for NGC 4736.  
\end{table*}

\begin{figure*}
\ \hspace{0.5cm} \
\parbox{10cm}{
\psfig{file=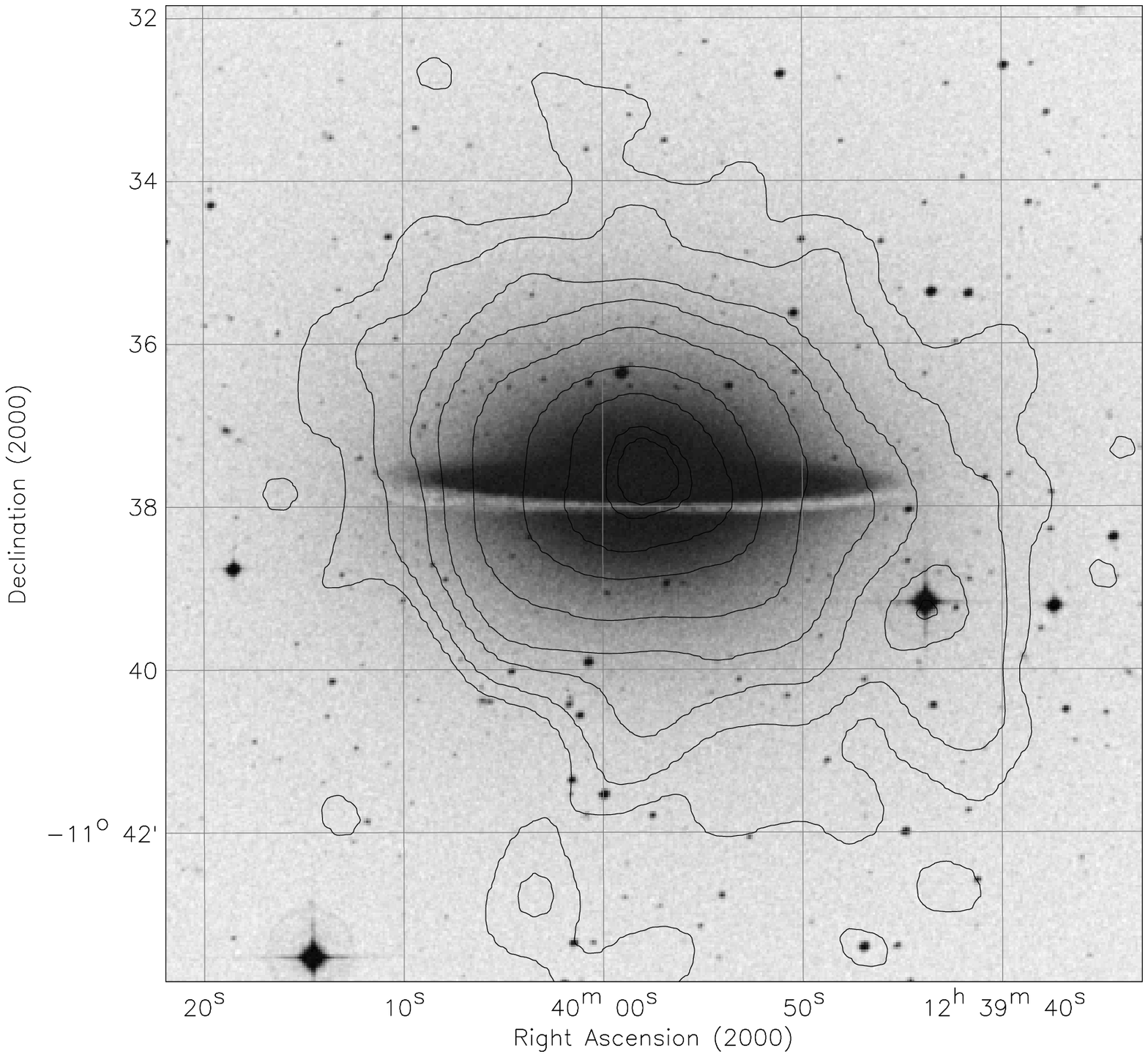,width=8.3cm,height=8.3cm,angle=0} }
\ \hspace{-0.1cm} \
\parbox{10cm}{
\psfig{file=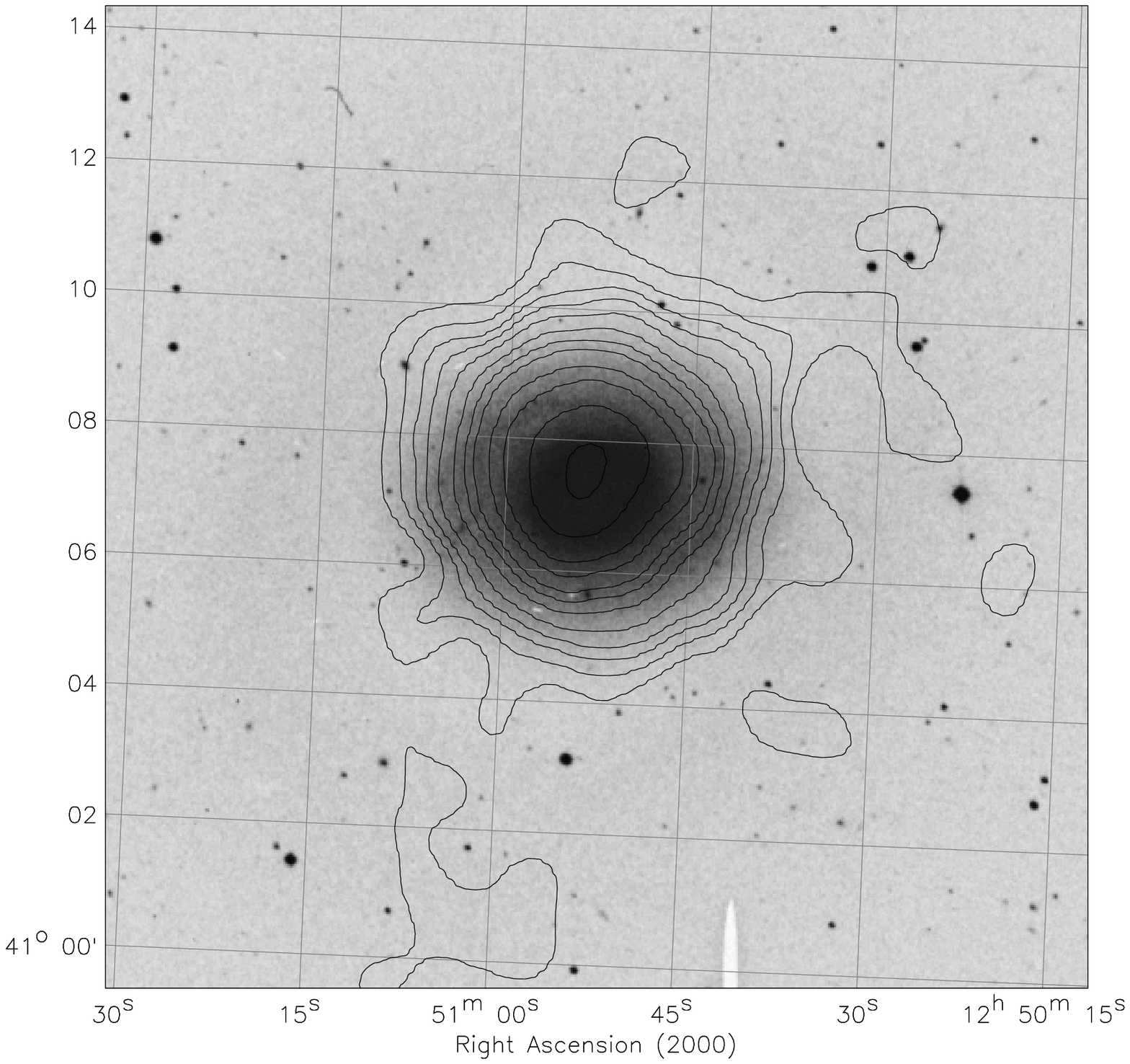,width=8.3cm,height=8.3cm,angle=0} }
\caption[]{The MECS image of Sombrero (left) and NGC 4736 (right)
in the 2--10 keV band,
smoothed with a gaussian of $\sigma =24^{\prime\prime}$. Contour
levels represent the 8, 10, 15, 20, 30, 50, 70, 90 and 95\% of the
peak intensity (left), and 6, 8, 10, 12, 15, 20, 25, 30, 40, 50, 70
and 95\% of it (right).
The contours are overlaid on the DSS image of the Space Telescope Science
Institute. }
\end{figure*}

\begin{figure*}
\ \hspace{0.3cm} \
\parbox{10cm}{
\psfig{file=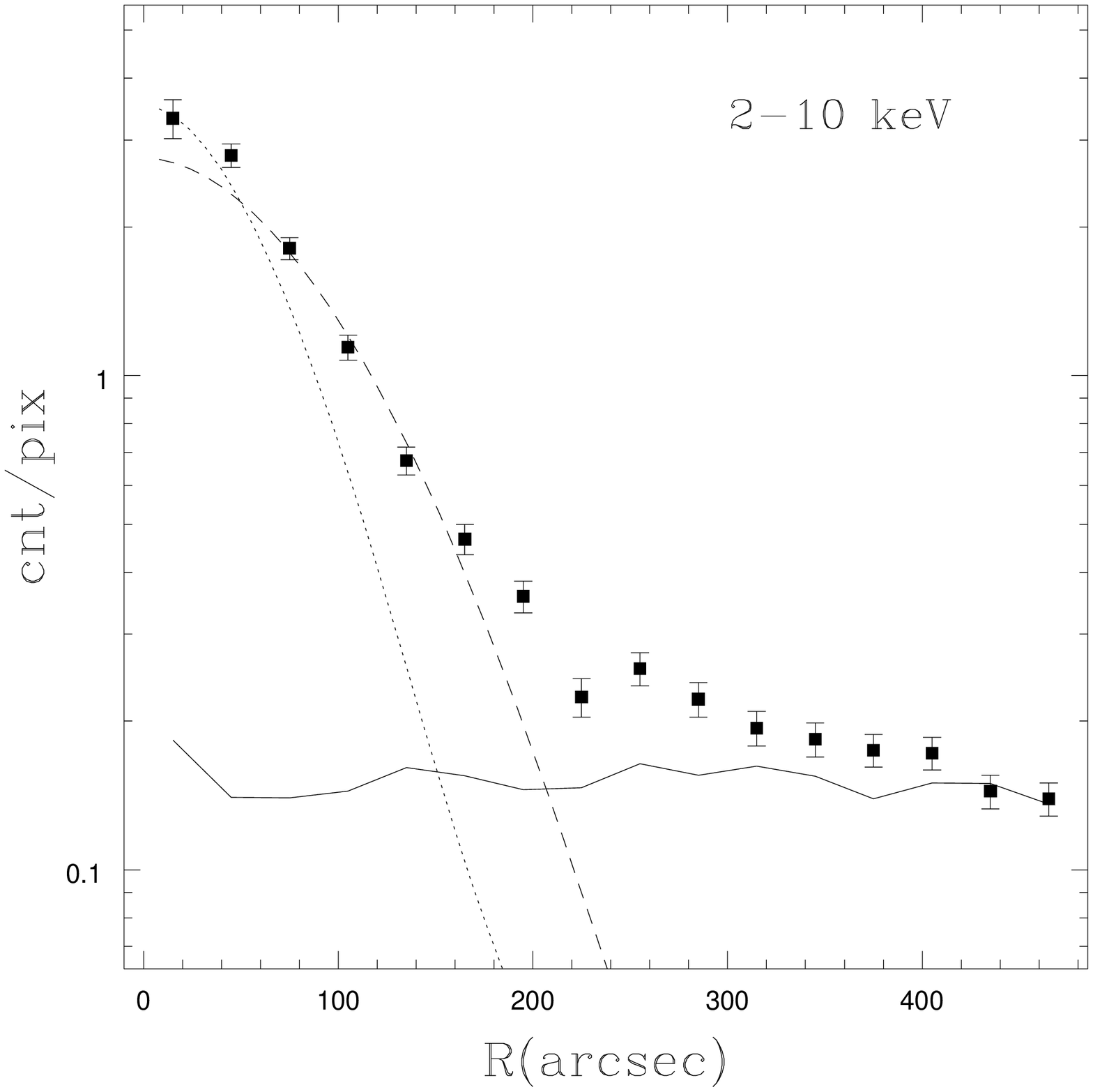,width=8cm,height=7cm,angle=0} }
\ \hspace{0.4cm} \
\parbox{10cm}{
\psfig{file=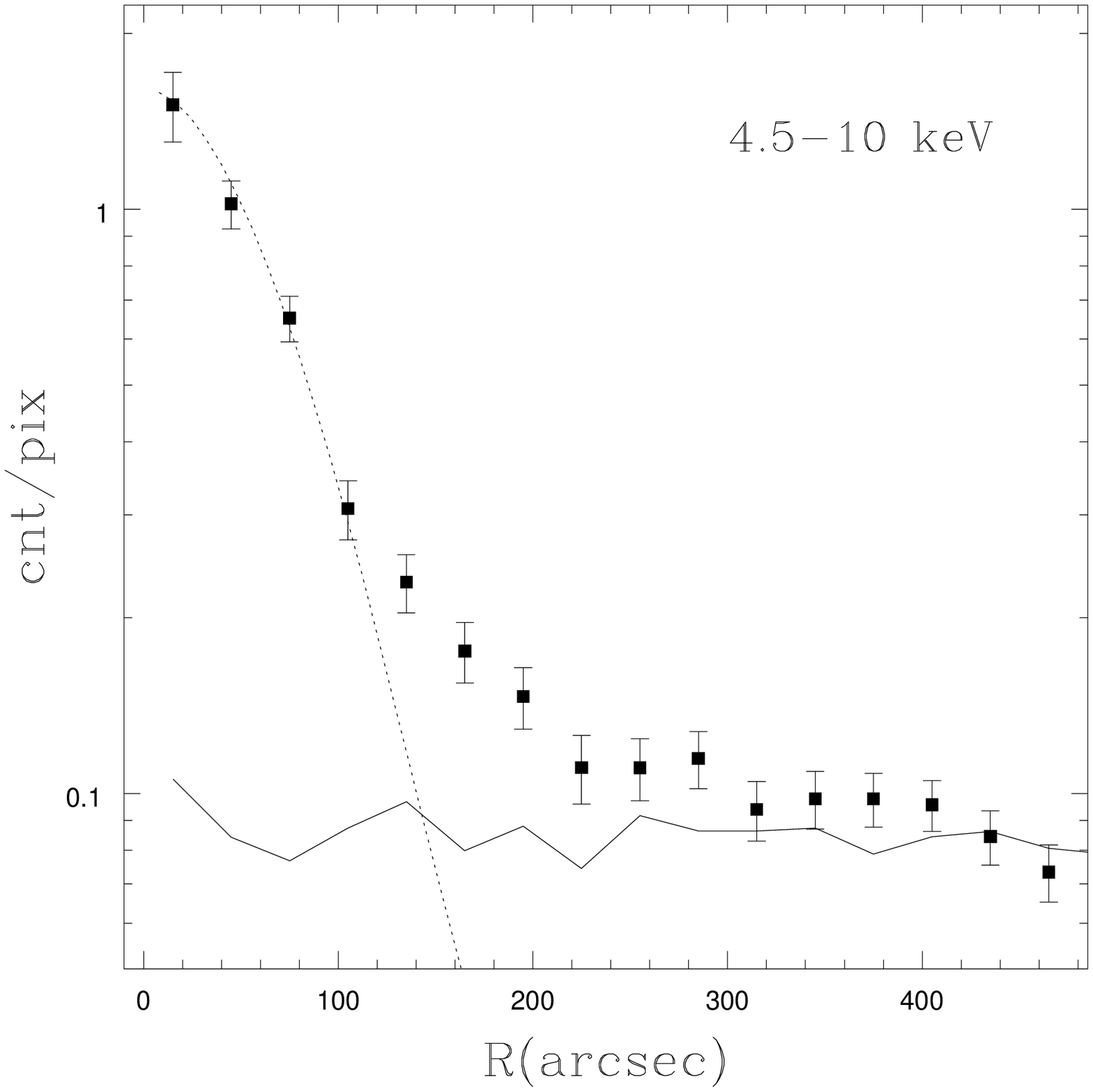,width=8cm,height=7cm,angle=0} }
\caption[]{The MECS 2--10 and 4.5--10 keV radial profiles of the total 
(squares) and background
(solid line) emission from Sombrero.  Also
shown is the PSF profile of 2 (dashed line) and 5 keV
(dotted line) photons.  A circular region of $32^{\prime\prime}$ radius
centered on the G0 star was omitted.}
\end{figure*}

\begin{figure*}
\ \hspace{0.3cm} \
\parbox{10cm}{
\psfig{file=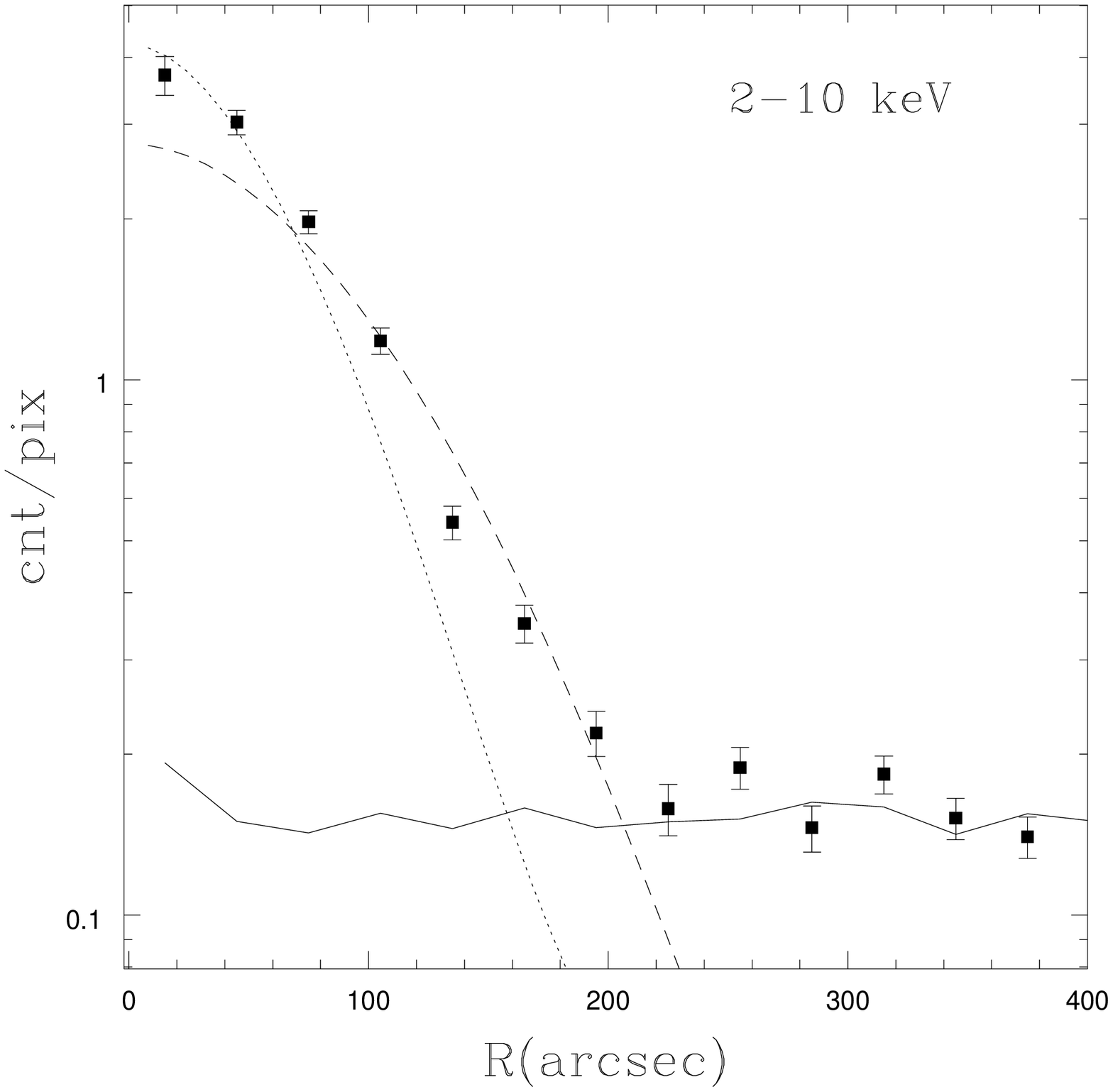,width=8cm,height=7cm,angle=0} }
\ \hspace{0.4cm} \
\parbox{10cm}{
\psfig{file=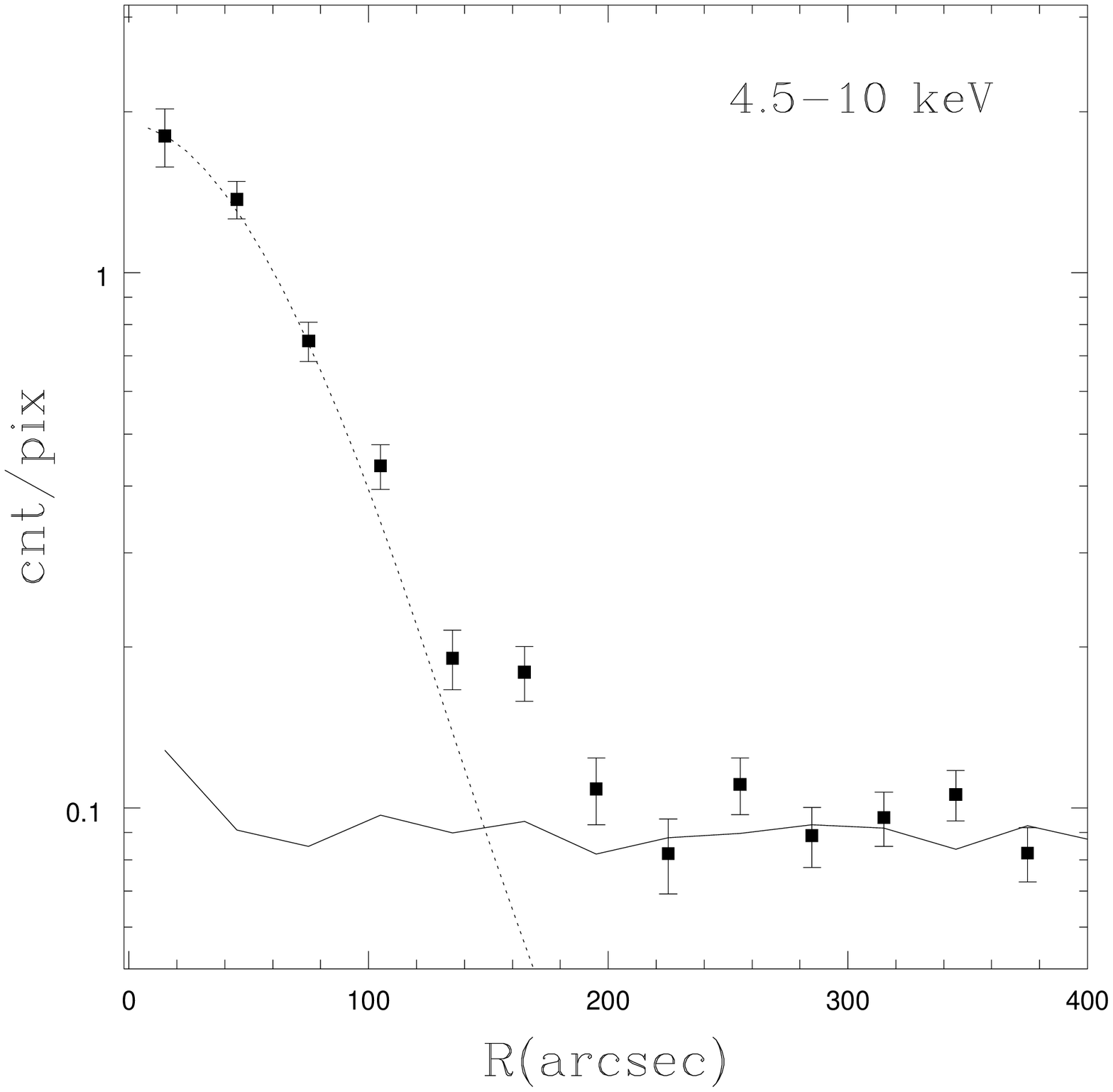,width=8cm,height=7cm,angle=0} }
\caption[]{The MECS 2--10 and 4.5--10 keV profiles of the total (squares) and background
(solid line) emission from NGC 4736. Dashed and dotted lines
represent the 2 and 5 keV PSF profiles.}
\end{figure*}

$BeppoSAX$ was pointed to the optical centers of the galaxies, on
 which the X-ray emission is peaked (Fig. 1) within the accuracy with
 which positions are given by the satellite (Boella et al. 1997). In
 the MECS image of Sombrero there is a prominent source at RA=12 39
 45.2, Dec=--11 38 49.6; we identify it with a bright 9.7 mag G0 star
 (HD 110086).  The absence of this source in the harder 4.5--10 keV
 image indicates its soft nature and supports its identification. We
 determined the spatial extent of the galactic emission using the MECS
 data\footnote{The PSF of the MECS includes 80\% of photons of
 energies $\geq 1.5$ keV within a radius of $2^{\prime}\hskip
 -0.1truecm .7$ (Boella et al. 1997). The LECS PSF is broader than
 this below 1 keV and similar above 2 keV.} from azimuthally averaged
 brightness profiles in concentric annuli centered on the X-ray
 centroids. The background profile in the same detector region was
 estimated from event files accumulated on different pointings of
 empty fields\footnote{These files are those released in Nov. 1998 for
 the MECS and Dec. 1999 for the LECS. A comparison of the background
 emission estimated from the field of the galaxies (in a region far
 from sources) with that estimated from the blank fields (in the same
 region, in detector coordinates) shows that the two estimates agree
 within 5\% for the MECS and 1\% for the LECS.}.  Figs. 2 and 3 show
 that the source emission becomes comparable to the background level
 at radii of $\sim 5-6^{\prime}$ for Sombrero and $\sim 3-4^{\prime}$
 for NGC 4736.  The hard emission is clearly extended in Sombrero
 relative to the instrumental PSF, while a less significant excess is
 seen in NGC 4736 above 4 keV. In order to reassure ourselves that the angular
 extension is not produced by the presence of the star in Sombrero, we
 derived radial profiles in four azimuthal sectors positioned along
 the north-south and east-west axes. The same conclusions as before
 are drawn.

The background-subtracted PDS count rate toward the Sombrero galaxy is
$(4.32\pm 1.35)\times 10^{-2}$ counts s$^{-1}$ in the 13--30 keV
band. This corresponds to a $\gsim 3\sigma$ detection. We have checked for
any possible contaminants in a region of 2 degrees radius around
Sombrero using the $ASCA$ SIS and GIS and $HEAO-1$ Source Catalogs; we
found no catalogued and bright hard X-ray sources.  Of course it is
 possible that both faint unknown hard X-ray sources and small
systematic errors in the background subtraction (Guainazzi \&
Matteuzzi 1997) contribute to the observed PDS count rate. For this
reason we consider this detection tentative. It must be confirmed by
hard X-ray imaging observations.
NGC 4736 was not detected by the PDS instrument; the net
13--30 keV count rate is $(1.8\pm 1.3)\times 10^{-2}$ counts s$^{-1}$.

\begin{table}
\caption[]{$Chandra$ ACIS-S observation Log}
\begin{flushleft}
\begin{tabular}{lllllllllllll}
\noalign{\smallskip}
\hline
\noalign{\smallskip}
Galaxy & \hskip-0.3truecm Obs ID & Date  & Exp.$^a$  &  Chip &
Sub.$^b$\\
\noalign{\smallskip}
\hline
\noalign{\smallskip}
Sombrero & 407  & 1999 Dec 12  & 1.766 & 7 (S3)& 1/8 \\
NGC 4736  & 808  & 2000 May 13  & 47.37 & 7 (S3)& 1/4 \\
\noalign{\smallskip}
\hline
\end{tabular} 
\end{flushleft}
$^a$ On-source net exposure time in ks.
 
$^b$ Subarray mode used to reduce pile-up.
\end{table}

\begin{figure}
\parbox{10cm}{
\psfig{file=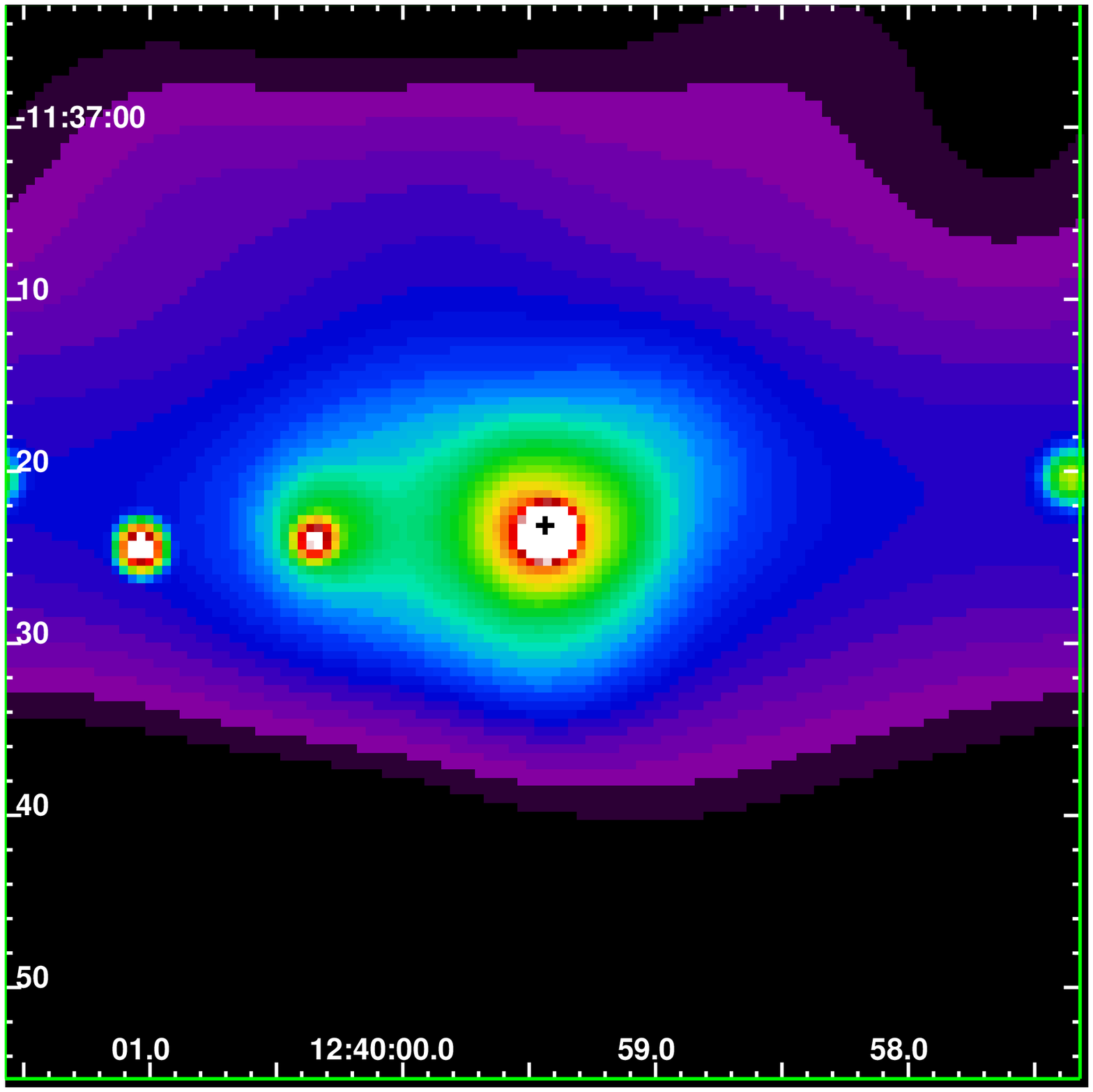,width=8.5cm,height=8.5cm,angle=0} }
\ \hspace{-0.5cm} \
\caption[]{The central $1^{\prime}\times 1^{\prime}$ portion of the
ACIS-S image of Sombrero in the 0.3--7 keV  band,
adaptively smoothed with gaussians of $\sigma$ ranging from
$0^{\prime\prime}\hskip -0.1truecm .5$ to $64^{\prime\prime}$. 
An unresolved radio source detected with the VLA at
8.4 GHz (Thean et al. 2000) is shown with a cross. }
\end{figure}

\subsection{$Chandra$ observations}

The journal of the $Chandra$ ACIS-S observations (Weisskopf, O'Dell \&
van Speybroeck 1996) is given in Table 3. We used
Level 2 reprocessed archived event files produced at the $Chandra$
X-ray Center (CXC).  The 0.3--7 keV images of the central region of
the pointed fields are shown in Figs. 4 and 5, where the data were
adaptively smoothed with the CXC CIAO tool csmooth. The positions of
two compact radio sources detected with the VLA in the nuclear regions
 are also marked; absolute positional accuracy for
these sources are $<0^{\prime\prime}\hskip -0.1truecm .5$ for NGC 4736
and $0^{\prime\prime}\hskip -0.1truecm .25$ for Sombrero, while the
$Chandra$ aspect solution is currently limited by systematics to $\sim
0^{\prime\prime}\hskip -0.1truecm .5$ accuracy (Aldcroft et
al. 2000).

Both the image (Fig. 4) and the radial profiles (Fig. 6) of Sombrero
indicate the presence of a bright and hard central point source,
coincident with the radio nuclear source, plus extended emission. This
becomes clearly visible at radii of
$2^{\prime\prime}-3^{\prime\prime}$ in the 0.3--7 keV band and at
$\sim 6^{\prime\prime}$ in the 1--7 keV band (from the comparison with
the $Chandra$ PSF, Fig. 6).  Therefore soft emission, most likely
coming from hot gas and/or soft binaries (see Sect. 5.1), gives a 
non-negligible contribution down to the very central region. The present
data do not allow a reliable assessment of its the shape and level due
to the very short exposure time.

\begin{figure}
\parbox{10cm}{
\psfig{file=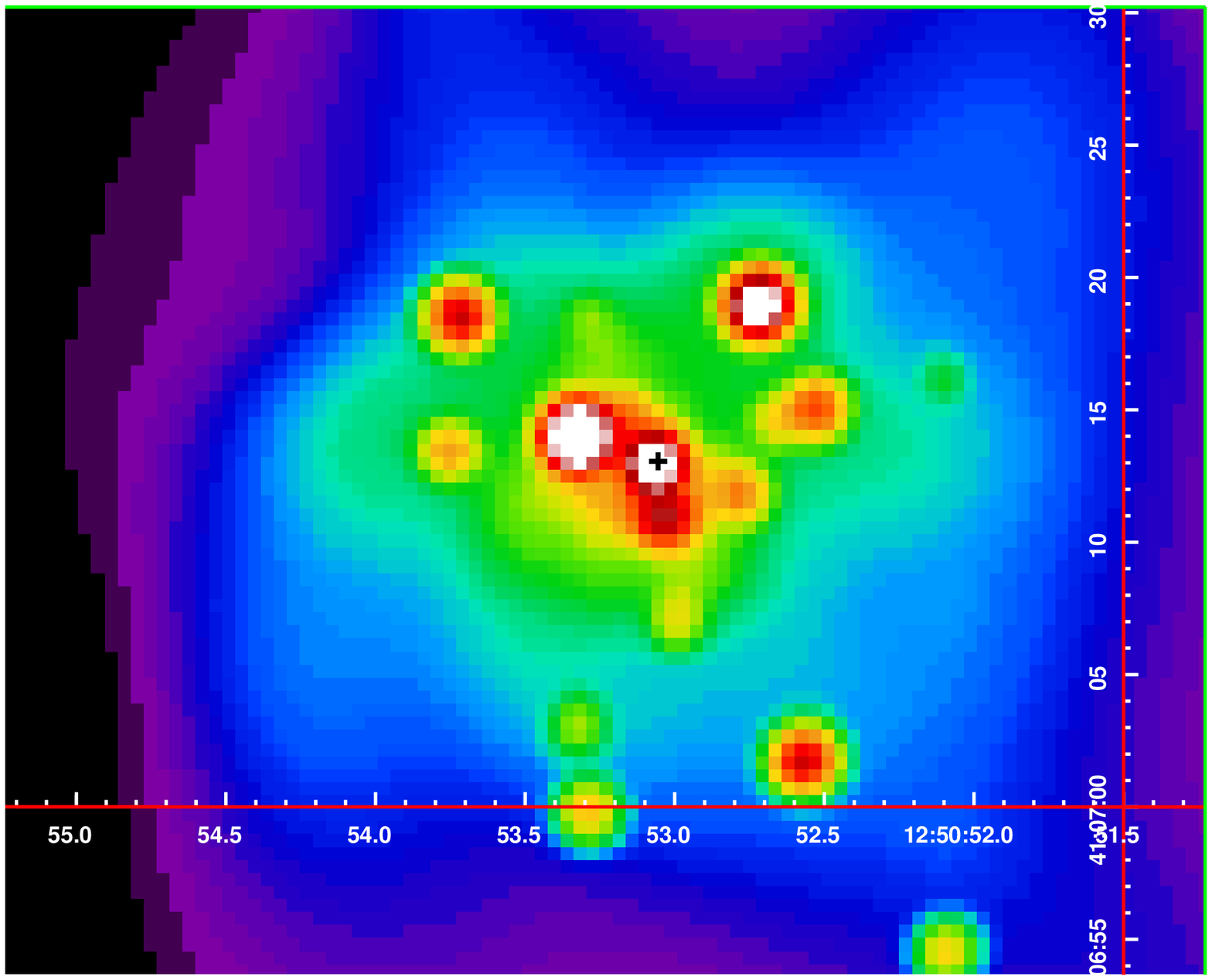,width=8.5cm,height=7.5cm,angle=0} }
\ \hspace{-0.5cm} \
\caption[]{The central $0^{\prime}\hskip -0.1truecm .5\times
0^{\prime}\hskip -0.1truecm .5$ portion of the ACIS-S image
of NGC 4736 in the 0.3--7 keV band, adaptively smoothed with
gaussians of $\sigma$ ranging from $0^{\prime\prime}\hskip -0.1truecm .5$ to
$64^{\prime\prime}$. The compact nuclear source detected
 with the VLA (Turner \& Ho 1994) is shown with a
cross. The bright source north-east of this X-ray/radio source
is the brightest of the field.}
\end{figure}

\begin{figure*}
\parbox{10cm}{
\psfig{file=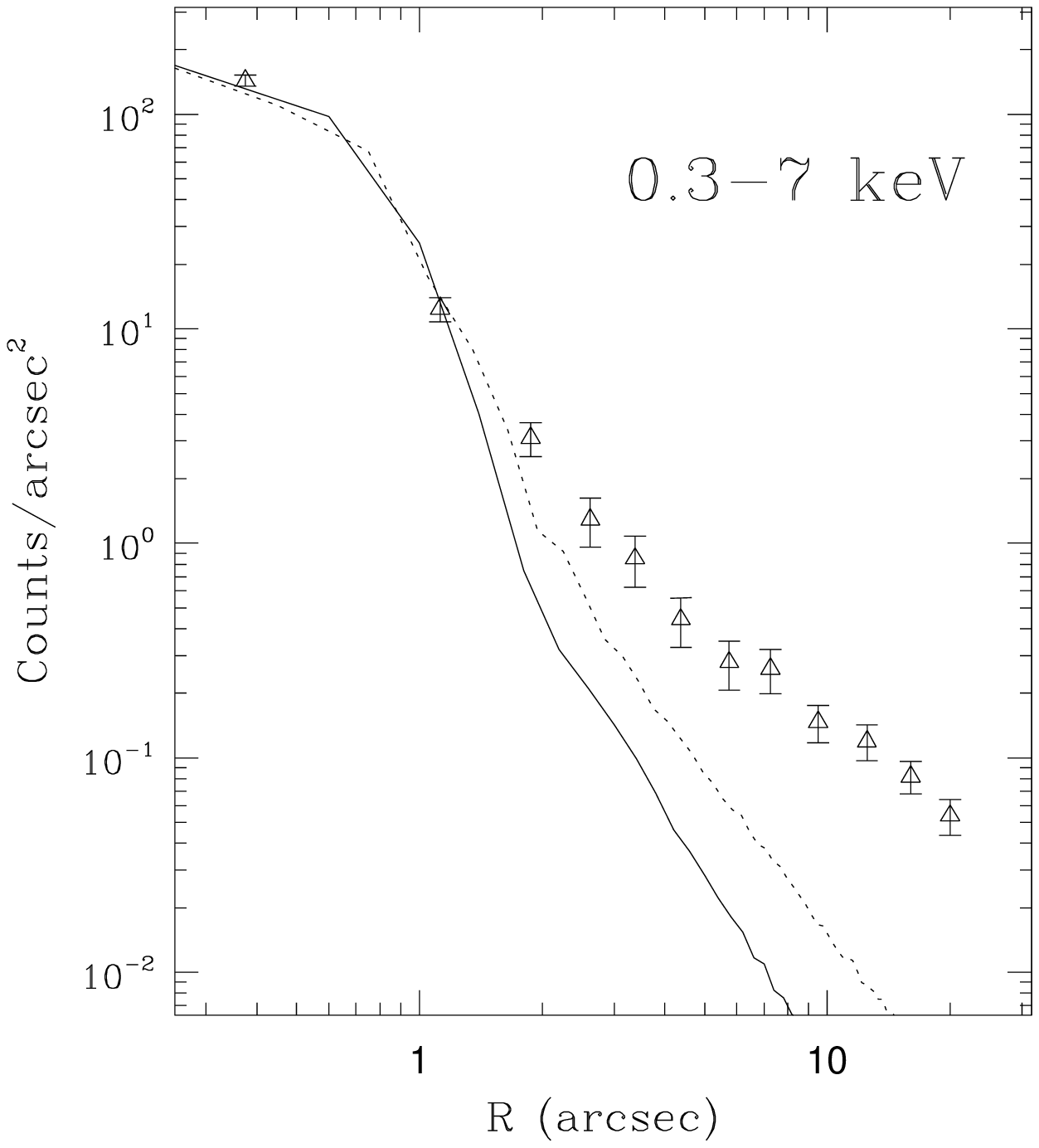,width=12.5cm,height=10.5cm,angle=0} }
\ \hspace{-3.6cm} \
\parbox{10cm}{
\psfig{file=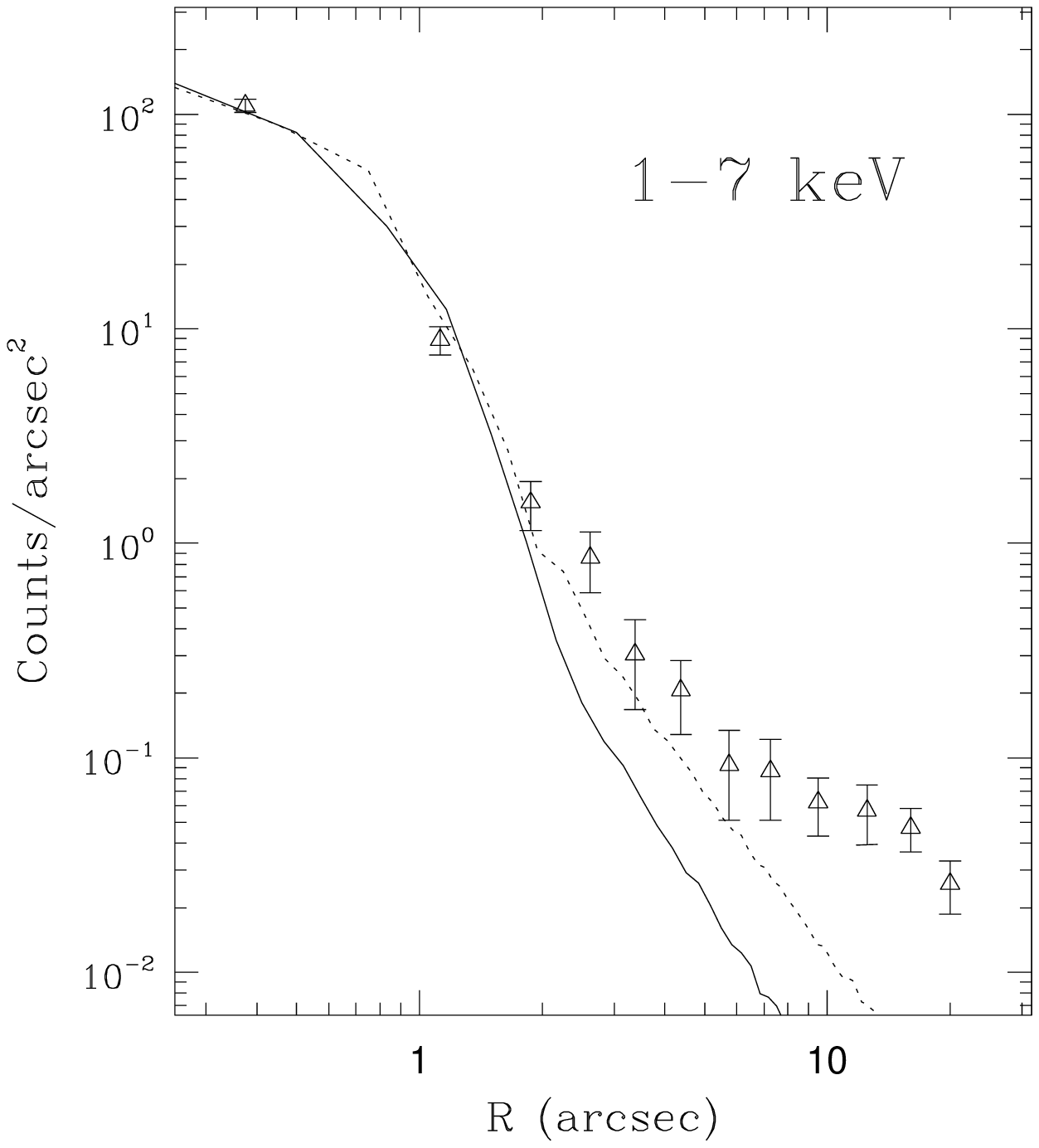,width=12.5cm,height=10.5cm,angle=0} }
\caption[]{ The ACIS-S radial profile of the center of Sombrero
(triangles) in the 0.3--7 keV and 1--7 keV bands, together with the
PSF profiles for 5 keV (dotted line), 0.8 keV (solid line, left panel)
and 1 keV (solid line, right panel). The PSF was estimated from
the standard $Chandra$ PSF library files for the source location in
the telescope field of view. The bright source $\sim
20^{\prime\prime}$ east of the nucleus (Fig. 4) was removed from
the calculation.}
\end{figure*}

At variance with Sombrero, various point sources are visible at the
center of NGC 4736, embedded within a bright diffuse emission
region. Moreover, the compact nuclear radio source does not coincide
with the brightest X-ray source, but with a much fainter one (Fig. 5).
The surface brightness profiles in the same physical region (the
central $\sim 1$ kpc) of Sombrero and NGC 4736 are compared in Fig. 7.
This comparison is aimed at showing the two galaxies as if they were
at the same distance from us, in the hard band.  Before
calculating its profile, we blurred the image of NGC 4736 to a
lower angular resolution (to compensate for the smaller distance) and we
 considered only data corresponding to a net exposure time equal
to that for Sombrero (extracted from the center of the
exposure interval).  Fig. 7 reveals that the profile of NGC 4736 is
less smooth than that of Sombrero, due to the presence of hard and bright
individual sources, and lacks a dominating central point
source.

\begin{figure}
\ \hspace{-1.cm} \
\parbox{10cm}{
\psfig{file=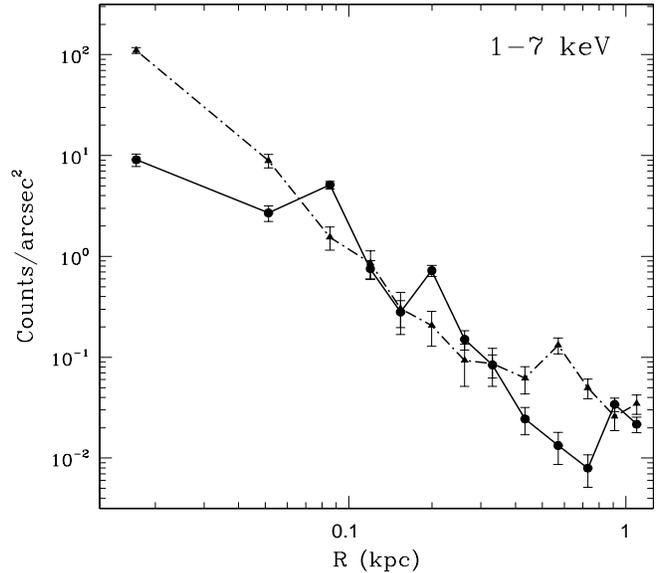,width=11.5cm,height=11cm,angle=0} }
\caption[]{The central 1--7 keV brightness profiles of Sombrero
(dot-dashed line) and NGC 4736 (solid line), from ACIS-S data. The
Sombrero profile is the same as in Fig. 6, except that no
source was removed from the
calculation. The NGC 4736 profile, centered on the radio nucleus,
comes from an image adapted from the original one, as detailed at the 
end of Sect. 2.2.}
\end{figure}

\section {Spectral analysis}

We investigated the spectral properties of the entire galaxies, using
$BeppoSAX$ data, and of the central point sources, using $Chandra$
data.  With the latter we also studied the overall spectral shape of
the largest central portion of the galaxies that was observed (more
details below) {\it just to check the consistency with the global
$BeppoSAX$ results}. 
The LECS and MECS background spectra were estimated from the same
blank fields used in Sect. 2.1; RMF and ARF files released in
September 1997 were used. The analysis of the LECS and MECS spectra
was done over 0.1--4 keV and 1.7--10 keV respectively.
We fitted the models simultaneously to the LECS and MECS data, and
then we considered the PDS ones in the case of Sombrero.  In the fitting
a normalization constant was introduced to allow for known differences
in the absolute cross-calibrations between the detectors (Fiore et
al. 1999). For the ACIS-S data, RMF and ARF files for the source and
background spectra were created using the script ``psextract'' in the
CIAO 2.1.2 software and the $Chandra$ CALDB version 2.6.  We derived
the background locally as detailed below.  Finally, source counts were
rebinned to obtain at least 20 counts per energy channel.  Spectral
fits were performed using XSPEC (version 11.0.1).  Quoted errors on
the fit parameters or upper limits refer to the 90\% confidence
interval for one interesting parameter ($\Delta \chi^2=2.71$).

\subsection{Sombrero}

\subsubsection{Analysis of $BeppoSAX$ data}
 
The spectral counts, extracted from a circle of $6^{\prime}$ radius
(Sect. 2.1), are well fitted by a simple power law of $\Gamma\sim 1.8$
with $N_H$ consistent with the Galactic value (Fig. 8, Table 4).  The
PDS data fall well above this model (in Fig. 8 the normalization
constant between the MECS and the PDS fluxes was fixed at its
canonical value).  We could not reproduce this excess by adding a
highly absorbed power law.  Since there are no catalogued bright hard
sources in the PDS field of view (Sect. 2.1), nor is any visible in
the MECS,  an unknown source could be producing the
excess. Its 13--30 keV flux would be $> 2.5\times 10^{-12}$ erg cm$^{-2}$
s$^{-1}$ when extrapolating the best fit MECS power law into the PDS
energy range and comparing the expected with the observed count rate
(considering also the PDS response outside the MECS field of view).

The addition of a soft thermal component (a MEKAL model) improves the
fit quality, but is significant at 95\% confidence level only,
according to the F-test.  Its best fit temperature is $kT\sim 0.3$
keV, its abundance is not constrained, and its presence requires an
intrinsic absorption to the power law component of
$N_{H,intr}=3.1\times 10^{21}$ \cm2\sp at the best fit (Table 4).

We also studied the MECS spectrum extracted from a circle of
$3^{\prime}$ radius to investigate the hard spectral properties of an
emission region that is more dominated by the nuclear emission. In
this case an emission line is found at $6.50^{+0.20}_ {-0.21}$ keV of
equivalent width $291_{-226}^{+205}$ eV (Fig. 9). The evidence for it
is marginal, though, since it is statistically significant at
93\% confidence only, according to the F-test. 

\begin{table}
\caption[]{Sombrero -- Results of the LECS and MECS spectral analysis.}
\begin{flushleft}
\begin{tabular}{  l l l l l l l l l l }
\noalign{\smallskip}
\hline 
\hline
\noalign{\smallskip}
Power law model:             &                       \\
\hline
\noalign{\smallskip}
$N_H $ ($10^{20}$ cm$^{-2}$) & 1.2 (0--5.8) \\
$\Gamma$                     & 1.85 (1.74--1.97) \\
Flux ($10^{-12}$ erg cm$^{-2}$ s$^{-1}$) & 1.5, 2.0 \\
Lum  ($10^{40}$  erg s$^{-1}$)           & 2.5, 2.1 \\
$\chi^2/\nu $ & 103/113     \\
\noalign{\smallskip}
\hline
\noalign{\smallskip}
Power law + Mekal model ($Z=0.5Z_{\odot}$):             &                   \\
\hline
\noalign{\smallskip}
$N_H $ ($10^{20}$ cm$^{-2}$) & 31 (2--70) \\
$\Gamma$                     & 1.95 (1.87--2.15) \\
Flux ($10^{-12}$ erg cm$^{-2}$ s$^{-1}$) & 0.7, 2.0 \\
Lum  ($10^{40}$  erg s$^{-1}$)           & 3.1, 2.1 \\
$kT$(keV)                    & 0.3 (0.2--0.5) \\
Flux ($10^{-12}$ erg cm$^{-2}$ s$^{-1}$) & 0.6, 0.  \\
Lum  ($10^{40}$  erg s$^{-1}$)           & 0.9, 0. \\
$\chi^2/\nu $   & 98/111     \\
\noalign{\smallskip}
\hline
\hline
\noalign{\smallskip}
\end{tabular} 
\end{flushleft}

$N_H$ is the column density of neutral hydrogen in addition to $N_{H,Gal}$
given in Table 1. Values between parentheses give the 90\% confidence 
interval.
$\nu $ is the number of degrees of freedom of the fit. 
Observed fluxes and
intrinsic luminosities are given in the 0.1--2 and 2--10 keV
bands.  

\end{table}

\subsubsection{Analysis of $Chandra$ data}

We studied the ACIS-S spectra of a central region of
$2^{\prime\prime}$ radius, where the emission from the central point
source dominates (Fig. 6), and from the largest accessible region of
the galaxy (this observation was performed in subarray mode), i.e., a
circle of $0.^{\prime}5$ radius. The background of the first spectrum
was derived locally, from an annulus of inner and outer radii of
$5^{\prime\prime}$ and $9^{\prime\prime}$. That of the $0.^{\prime}5$
spectrum comes from a circle of $25^{\prime\prime}$ radius located as
far as possible from the center of the galaxy (in the north direction)
but still within the S3 chip. The spectral fits were performed over
0.5--5 keV, because not enough counts were detected above 5 keV.  The
spectra are well described by simple power laws, whose $\Gamma \sim
1.5$ and 1.6 at the best fit, respectively for the small and the
large extraction region (Table 5).  The uncertainties on the
spectral parameters are large, due to the short exposure time
(Fig. 10, left panel).  Taking the extreme values on the 90\%
confidence contour in this figure, the 2--10 keV observed
flux of the central spectrum ranges from $1.1\times 10^{-12}$ to
$2.2\times 10^{-12}$ \ecs\sp [and its L(2--10 keV)=$(1.2-2.3)\times
10^{40}$ \es].  In the 0.5--5 keV band, the $2^{\prime\prime}$ radius
region contributes $\sim 60$\% of the counts from the $0.^{\prime}5$
one; in the 2--10 keV band it contributes $\sim 70$\% of them. At
least one clear difference between the two spectra is found (Fig. 10,
left panel): the central one requires an absorption in excess of the
Galactic value, while the spectrum from $0.^{\prime}5$ does not.  This
result is in agreement with the absence of intrinsic absorption found
from $BeppoSAX$ data.  Fig. 10 (right panel) also shows some overlap
between the spectral results obtained from the LECS+MECS data for the
whole galaxy and those obtained for the $0.^{\prime}5$ radius region
from ACIS-S data. The $\Gamma$ value given by $BeppoSAX$ tends to be
larger, though, as if the galactic emission were softer than that
of the central region. This is in accordance with the
steepening of the power law slope found by FJ from $ROSAT$ PSPC data
when increasing the extraction radius (Sect. 1.1).  In the
$0.^{\prime}5$ ACIS-S spectrum there is no significant evidence for a
soft thermal component, however the data uncertainties are large.

\begin{figure*}
\parbox{10cm}{
\psfig{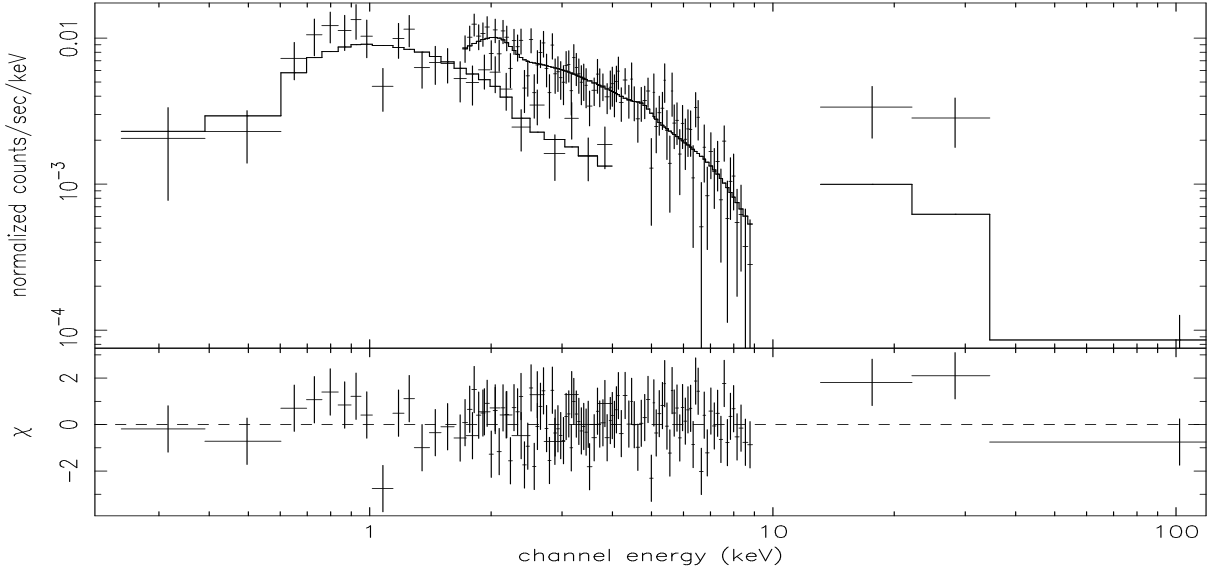} }
\ \hspace{-0.5cm} \
\caption[]{The LECS, MECS and PDS spectra of Sombrero together with
their modeling with a power law (Sect. 3.1.1, Table 4). Folded spectra
and residuals are shown.}
\end{figure*}

\subsection{NGC 4736}

\subsubsection{Analysis of $BeppoSAX$ data}

Spectral counts were extracted from a circle of $4^{\prime}$ radius
(Sect. 2.1).  Their fit with a simple power law is of poorer quality
than for Sombrero (Table 6), mostly because this model leaves an
excess of emission around 1 keV (Fig. 11).  The addition of a thermal
component of $kT\sim 1$ keV and extremely low abundance improves
significantly the quality of the fit, but this model is not
realistic. Moreover, this soft component largely dominates the
observed emission below 2 keV, because the power law becomes highly
absorbed ($N_{H,intr}>2\times10^{22}$\cm2), in contrast with what
found from $ROSAT$ data (Sect. 1.1).  The abundance is not constrained
by the LECS data, though, and if we fix it at any $Z\gsim
0.1Z_{\odot}$, any intrinsic $N_H$ to the power law is $<1.4\times
10^{20}$\cm2, while the addition of the thermal component still gives
a significant improvement ($Z=Z_{\odot}$ in Table 6; if $Z=0.1
Z_{\odot}$, the values of $kT$ and $\Gamma$ are still
within the uncertainties in Table 6).  The addition of a second soft
thermal component does not improve the fit quality further.  No
emission lines at high energies are detected; the equivalent width of
any Fe-K emission at 6.4 or 6.7 keV is $<380$ eV and $<420$ eV
respectively.

\begin{table*}
\caption[]{Spectral results from ACIS-S data for central point sources.}
\begin{flushleft}
\begin{tabular}{  l l l l l l l l l l }
\hline
\hline 
\noalign{\smallskip}
Power law model:             & Sombrero nucleus & NGC 4736 radio nucleus$^a$ & NGC 4736 brightest source\\
\hline
\noalign{\smallskip}
$N_H $ ($10^{20}$ cm$^{-2}$) & 17 (8--28)    & 27 (7-66)          & $<1.5$ \\
$\Gamma$                & 1.5 (1.2--1.9)& 1.9 (1.8--2.2) & 1.22 (1.19--1.29) \\
Flux ($10^{-12}$ erg cm$^{-2}$ s$^{-1}$) & 1.6 &  0.2          & 0.7 \\
Lum  ($10^{39}$  erg s$^{-1}$)           & 17 &  0.6           & 3.0\\
$\chi^2/\nu$             & 0.68             &   0.98             & 0.89  \\
\noalign{\smallskip}
\hline\hline
\end{tabular} 
\end{flushleft}

The $N_H$ value is in addition to $N_{H,Gal}$ in Table 1.  Values
between parentheses give the 90\% confidence interval.  Observed
fluxes and intrinsic luminosities refer to the 2--10 keV band.

$^a$ A thermal component was added to the power law (Sect. 3.2.2).

\end{table*}

\begin{figure}
\parbox{10cm}{
\psfig{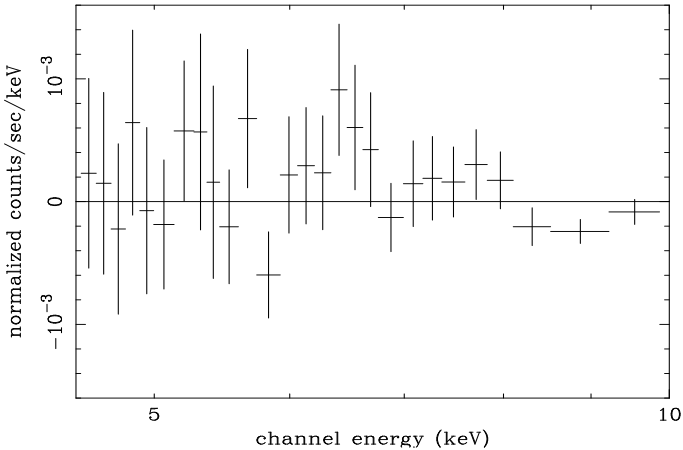} }
\ \hspace{-0.5cm} \
\caption[]{The residuals between the MECS spectrum of Sombrero extracted from 
the central $3^{\prime}$ and a simple power law
 model, around the 6.5 keV ``emission line'' (Sect. 3.1.1).}
\end{figure}

\begin{figure*}
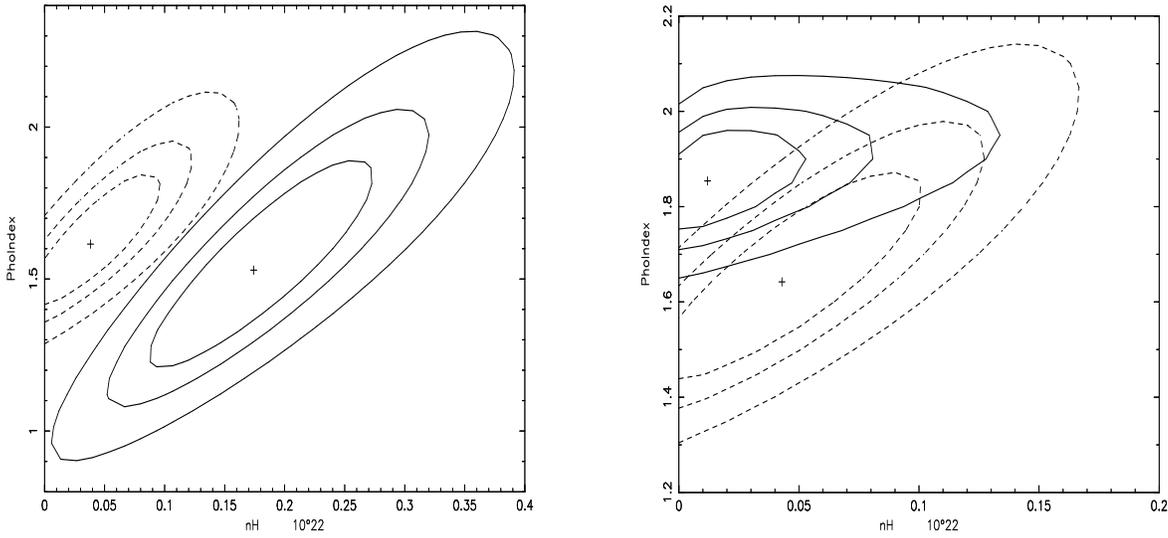

\parbox{10cm}{
\psfig{file=fig10a.ps,width=7cm,height=7cm,angle=-90} }
\ \hspace{1cm} \
\parbox{10cm}{
\psfig{file=fig10b.ps,width=7cm,height=7cm,angle=-90} }
\caption[]{The 68\%, 90\% and 99\% confidence contours for 
$\Gamma$ and $N_{H,intr}$
derived for Sombrero (Sects. 3.1.1 and 3.1.2).
Dashed contours refer to the ACIS-S spectrum extracted from
a $0.^{\prime}5$ radius; solid contours
refer to the ACIS-S spectrum from
a $2^{\prime\prime}$ radius in the left panel, and the LECS+MECS spectrum
in the right panel.}
\end{figure*}

\begin{figure*}
\parbox{10cm}{
\psfig{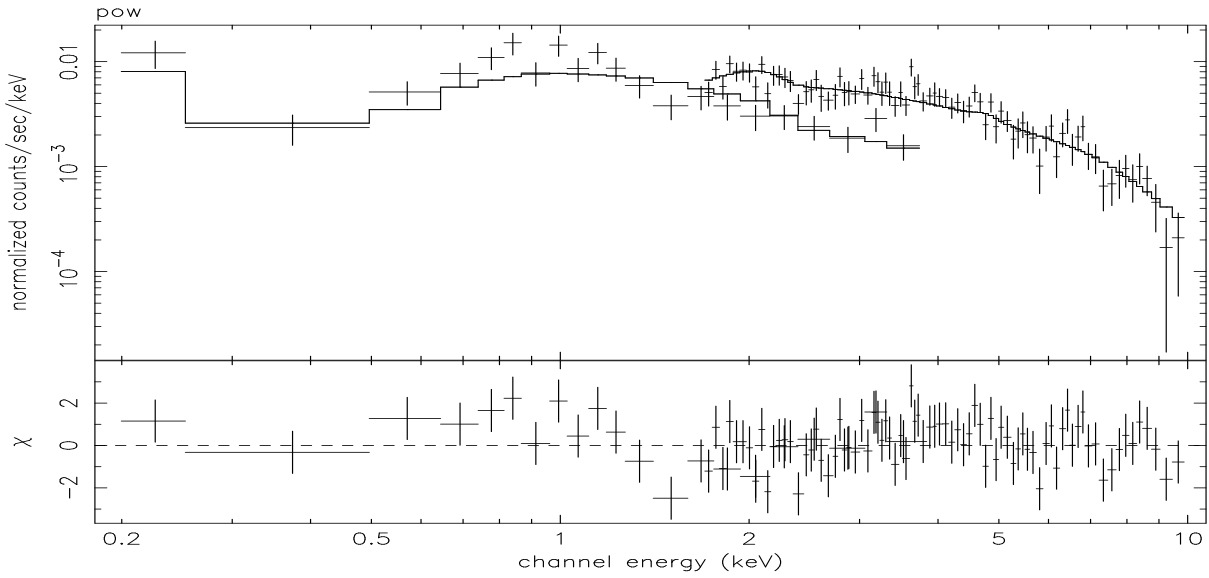} }
\ \hspace{-0.5cm} \
\caption[]{The LECS and MECS  spectra of NGC 4736 fitted
with a power law (Table 6). Folded spectra and residuals are shown.}
\end{figure*}

\subsubsection{Analysis of $Chandra$ data}

We studied the spectra of two point sources in the nuclear
region: the brightest one, which is also the brightest of the field,
and the one coincident with the nuclear radio source
(Fig. 5). We extracted counts from a circle of $1.^{\prime\prime}5$
radius and we used as background a circle of $5^{\prime\prime}.5$
radius, centered on the X-ray/radio point source, from which we
removed the point sources lying within.  We also studied the spectral
shape of the largest accessible portion of the galaxy, a circle of
$55^{\prime\prime}$ radius, for which the background was taken from a
circle of $50^{\prime\prime}$ radius, located as far as possible from
the center of NGC 4736 but still within the S3 chip.  The spectral fits
were performed over the 0.5--7 keV range.

Both point sources can be modeled by power law spectra, however with
significantly different parameters (Tab. 5): the brightest one has
$\Gamma =1.22$ without intrinsic absorption, the X-ray/radio point
source is softer and absorbed (Fig. 12). Moreover, the addition of a
soft thermal component gives a significant improvement of the fit of
this source. Since it is embedded in a diffuse emission region
(Fig. 5), we checked whether a ``pure'' source spectra could be
obtained by restricting further the extraction radius and by using
various alternative background estimates (i.e., obtained from a region
closer to the point source, or inserting it as a thermal component
with free normalization).  The fitting results always remained very
similar.  If we adopt the $1.^{\prime\prime}5$ source extraction
radius and a thermal model for the the background, then $\Gamma\sim
1.9$ for the power law component (Tab. 5), and $kT=0.65^{+0.05}_
{-0.06}$ keV, $Z=0.07^{+0.27}_{-0.01} Z_{\odot}$ for the thermal
component.  We also tried to reproduce the softer part of the spectrum
with a multicolor disk blackbody emission, in addition to a power law,
which is the usual modeling adopted for black hole binaries (Tanaka \&
Shibazaki 1996).  The quality of the fit improves with respect to that
with a simple power law, but is not as good as when a thermal
component is added.

The ``global'' galaxy spectrum requires at least two thermal
components plus a power law one (of $\Gamma=1.47^{+0.04}_{-0.05}$)
without intrinsic absorption.  The two temperatures are
$kT=0.23^{+0.13}_{-0.04}$ keV (with $Z=0.02 Z_{\odot}$ at the best
fit, and $Z<1.6 Z_{\odot}$) and $kT=0.57^{+0.01}_{-0.02}$ keV (with
$Z=0.8Z_{\odot}$ at the best fit, and $Z>0.2 Z_{\odot}$). When
constraining the abundances to be the same, their common value becomes
$Z=0.24^{+0.27}_{-0.08} Z_{\odot}$ (the other spectral parameters
remain practically unchanged).  Note that these ``global'' fit results
are only preliminary, awaiting for improved soft-band calibration of
ACIS-S.  The $BeppoSAX$ and the ``global'' $Chandra$ spectra give
consistent photon index values.  At lower energies, the higher
sensitivity and spectral resolution of ACIS-S can distinguish the
presence of two thermal components.

\begin{table}
\caption[]{NGC 4736 -- Results of the LECS and MECS  spectral 
analysis. }
\begin{flushleft}
\begin{tabular}{  l l l l l l l l l l }
\noalign{\smallskip}
\hline
\hline 
\noalign{\smallskip}
Power law model:             &                       \\
\hline
\noalign{\smallskip}
$N_H $ ($10^{20}$ cm$^{-2}$) & 0.0 ($<0.7$) \\
$\Gamma$                     & 1.69 (1.60--1.78) \\
Flux ($10^{-12}$ erg cm$^{-2}$ s$^{-1}$) & 1.3, 1.9 \\
Lum  ($10^{39}$  erg s$^{-1}$)           & 7.0, 7.7 \\
$\chi^2/\nu $  & 112/97    \\
\noalign{\smallskip}
\hline
\noalign{\smallskip}
Power law + Mekal model ($Z=Z_{\odot}$):      &                       \\
\hline
\noalign{\smallskip}
$N_H $ ($10^{20}$ cm$^{-2}$) & 0 ($<1.4$) \\
$\Gamma$                     & 1.63 (1.52--1.71) \\
Flux ($10^{-12}$ erg cm$^{-2}$ s$^{-1}$) & 1.2, 1.9  \\
Lum  ($10^{39}$  erg s$^{-1}$)           & 6.2, 7.9\\
$kT$(keV)                    & 0.3 (0.2--0.6) \\
Flux ($10^{-12}$ erg cm$^{-2}$ s$^{-1}$) & 0.4, 0. \\
Lum  ($10^{39}$  erg s$^{-1}$)           & 1.9, 0.  \\
$\chi^2/\nu $ & 99/95    \\
\noalign{\smallskip}
\hline
\hline
\end{tabular} 
\end{flushleft}

$N_H$ is the column density of neutral hydrogen in addition to
$N_{H,Gal}$ given in Table 1.  Values between parentheses give the
90\% confidence interval.  $\nu $ is the number of degrees of freedom
of the fit.  Observed fluxes and intrinsic luminosities are given in
the 0.1--2 and 2--10 keV bands.

\end{table}

\section {Variability}

Given the complexity of the sources considered, a variability study
with low angular resolution data is clearly limited.  For a comparison
with previous analyses, though, we use the $BeppoSAX$ data to study
the short and long term variability.

\begin{figure*}
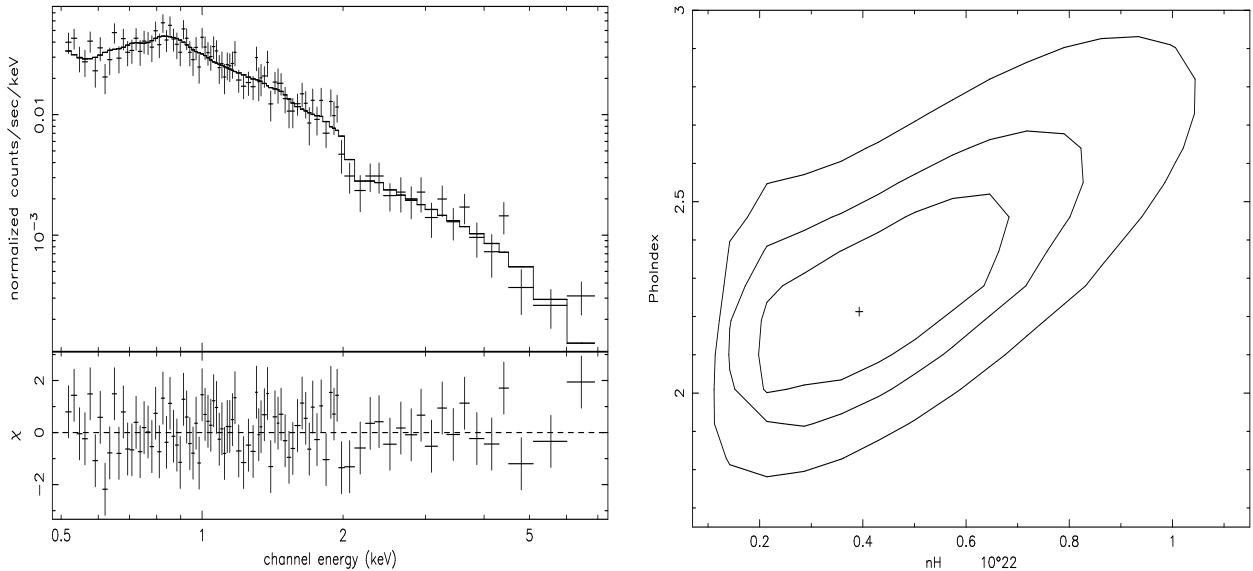

\parbox{10cm}{
\psfig{file=fig12a.ps,width=8cm,height=7.5cm,angle=-90} }
\ \hspace{0.1truecm} \
\parbox{10cm}{
\psfig{file=fig12b.ps,width=8cm,height=7.5cm,angle=-90} }
\caption[]{Left: the ACIS-S spectrum of the X-ray/radio point source
in NGC 4736, fitted with a power law plus a thermal component
(Sect. 3.2.2). Right: 68, 90 and 99\% confidence contours for the
parameters $N_{H,intr}$ and $\Gamma$ describing the model on the
left.}
\end{figure*}

$ROSAT$ HRI data showed that the nuclear source in Sombrero might be
time variable (FJ).  To investigate this point we extracted the 2--4
keV, 4--10 keV and 2--10 keV light curves for the MECS count rates of
the central $3^{\prime}$ radius circle.  Just a hint of variability
was found. A $\chi^2$ test on the 2--10 keV data, to quantify how the
observed count rate differs from a constant average emission, gives a
probability of chance occurrence of $\sim 80$\%.  So, the ``nuclear''
emission does not show a statistically significant variation on the
time scales sampled here. $ROSAT$ and $ASCA$ data
do not show evidence of significant short term temporal variability
for NGC 4736 (R99), and the same is found from the 2--10 keV MECS data.
In conclusion,
for both galaxies we can at least exclude variations of the 2--10 keV
flux larger than $\sim 50$\% on time scales of the order of one day.

The sources are stable also in the long term. $ASCA$ and $BeppoSAX$
 2--10 keV fluxes for Sombrero are consistent with each other (2.2 and
 2.0$\times 10^{-12}$ \ecs\sp respectively, pointings made during
 1994 and 2000). NGC 4736 showed roughly the same 2--10 keV
 flux during the $ASCA$ and $BeppoSAX$ pointings in the years 1995 and
 2000 (1.8 and 1.9 $\times 10^{-12}$ \ecs\sp respectively).
 R99 also found the hard continuum level unchanged between the $ROSAT$
 PSPC and $ASCA$ pointings (separated by four years).

\section{Summary and discussion}

\subsection{Sombrero}

The main findings derived here are:

\par\noindent 
1) the hard ($>2$ keV) MECS emission is
extended, consequently it has a non-negligible contribution from distributed
sources (e.g., XRBs).

\par\noindent 
2) a simple power law ($\Gamma\sim 1.8$) without absorption in excess
of the Galactic value reproduces well the LECS+MECS spectrum.
The addition of soft thermal emission ($kT\sim 0.3$ keV) improves 
the spectral fit, but is not strictly required.  

\par\noindent
3) at the galactic center a short ACIS-S exposure shows 
a single bright and hard source plus additional soft extended emission.
 
\par\noindent 
4) the ACIS-S spectra of the nucleus and of a central circle of
$0.^{\prime}5$ radius can both be described by simple power laws. The
nucleus requires absorption intrinsic to Sombrero, while the larger
region does not, in agreement with the $BeppoSAX$ result. 

\par\noindent 
5) an emission line consistent with fluorescence from cold iron is
marginally detected only in the MECS spectrum of the
central $3^{\prime}$ circle, so (if real) it could originate in the nucleus. 
The statistics of ACIS-S data above 5 keV are too low 
to investigate this point.

\par\noindent 
6) no significant short or long term flux variability is found. 

Below we discuss the global galaxy X-ray properties
 and then (Sect. 5.2) we consider the possibility of the
presence of a low luminosity AGN (LLAGN).

Soft thermal emission is indicated by the $BeppoSAX$ spectrum and by
 the ACIS-S 0.3--7 keV brightness profile. 
  Its origin is likely due to
 hot gas and/or stellar sources with soft emission distributed over
 the galaxy [see, e.g., those discussed in Pellegrini \& Fabbiano
 (1994), or black hole binaries (Tanaka \& Shibazaki 1996)].  The
 presence of some hot gas in this bulge-dominated galaxy is plausible.
 The temperature and luminosity of the soft component (Table 4) agree
 with those expected for the stellar mass losses that can be retained
 by potential wells of the shape and depth of the Sombrero galaxy
 (D'Ercole \& Ciotti 1998).  The central stellar velocity dispersion
 of the bulge ($\sigma=249$ km s$^{-1}$, McElroy 1995) corresponds to
 a central gas virial temperature of 0.37 keV, close to the global
 value found here ($kT= 0.3$ keV).  The diffuse emission
 estimated from the HRI data, including possible truly diffuse gas and
 unresolved sources, is L(0.1--2.4 keV)$\sim 1.2\times 10^{40}$ \es
 (FJ, rescaled for the distance adopted here), consistent with
L(0.1--2.4 keV)$\sim 0.9\times 10^{40}$ \es\sp derived from $BeppoSAX$ data
for the soft thermal emission.

The global galactic emission in the hard band is largely contaminated
by the nuclear source. From ACIS-S data (Sect. 3.1.2), this source
contributes $\gsim 55$\% of the MECS 2--10 keV galactic flux.  Given
that the latter is extended (Fig. 2), there must be some hard emission
in excess of the nuclear one, likely due to XRBs.  By subtracting the
ACIS-S nuclear emission from the $BeppoSAX$ spectrum, the hard XRB's
contribution is L(2--10 keV)$\lsim 9\times 10^{39}$\es. This upper
limit is very close to the XRB's contribution in Sombrero estimated
from the best fit $L_X-L_B$ relation (converted to the 2--10 keV band)
of normal spiral galaxies, based on $Einstein$ data [i.e., L(2--10
keV)$= 8.7\times 10^{39}$ \es; Fabbiano, Kim \& Trinchieri 1992].  It
is also in agreement with the estimate ($\sim 4\times 10^{39}$\es,
when converted to the 2--10 keV band) from the $L_X-L_B$ relation for
the collective emission of discrete non-nuclear X-ray sources in
spirals, derived recently from a large sample of nearby galaxies
observed with the $ROSAT$ HRI (Roberts \& Warwick 2000).

\subsection{The low luminosity AGN in Sombrero}

ACIS-S reveals a nuclear spike of  L(2--10 keV)=$(1.2-2.3)\times 10^{40}$\es,
 whose spectrum is well modeled by an
absorbed power law. The small amount of absorption 
[$N_{H,intr}=(0.8-2.8)\times 10^{21}$\cm2] could be
explained by intervening dust along the line of sight to the center.
An $HST$ V--I image, in fact, shows that the nucleus is partially
hidden by a dusty environment (Pogge et al. 2000), for which Emsellem
\& Ferruit (2000) derive $A_V$=0.5 mag (i.e., $N_{H,intr}=8.3\times
10^{20}$\cm2, for the Galactic extinction law).

Within the uncertainties, still large, the spectral characteristics
are consistent with both Seyfert galaxies (Nandra et al. 1997) and
XRBs.  Could a few bright XRBs be at the origin of the nuclear
emission and not be resolved by ACIS-S? This hypothesis requires the
presence of hard stellar sources in larger number and/or of higher
luminosity than in NGC 4736, within a radius of $\sim 100$ pc from the
center of Sombrero (e.g., Fig. 7).  This seems unlikely, given that it
would imply a recent starburst activity, at least as vigorous as in
NGC 4736 (Sects. 1.1 and 5.3), for which there is no evidence.
Therefore, the possibility of the presence of a LLAGN is more
viable. This also ties up with observational results outside the X-ray
band, i.e., with the indications for the presence of a central
supermassive black hole, of weak optical activity, of compact radio
and UV emissions, and also of a faint broad $H\alpha$ component
(Sect. 1.1).

What kind of physical mechanism could be at the origin of the nuclear
emission?  Its L(2--10 keV) derived here and the $H\alpha$ luminosity
estimate from the spectroscopic survey of Ho et al.  (1997) put this
nucleus very close to the extension down to low luminosities of the
$L_X-L_{H\alpha}$ correlation found for powerful Seyfert 1 nuclei and
quasars (Ward et al. 1988, Ho et al. 2001).  This indicates that the
optical emission line spectrum could be powered by photoionization
from the central continuum, as in high luminosity AGNs. This nucleus
could then behave like a downsized version of a bright AGN, a
possibility investigated more in detail for the LLAGN in NGC 4258 by
Fiore et al.  (2001).  Support for this interpretation could also come
from the possible presence of a $\sim 6.4$ keV emission line in the
MECS spectrum (Sect. 3.1.1), as often observed in Seyfert galaxies
(Nandra et al. 1997). Its equivalent width is similar to that of
Type 1 AGNs, while stronger lines are usually found in Type 2 AGNs,
consistent with the fact that large absorption is not required by our
spectral analysis.

A detailed study and modeling of the global spectral energy
 distribution (SED) of the nuclear emission is beyond the scope of
 this work. We just note that this SED differs from the canonical
 broadband continuum of bright AGNs in the lack of an ultraviolet
 excess (Nicholson et al. 1998), a property common to a
 few LLAGNs (Ho 1999, who also argues that it is not an artifact of
 dust extinction).  This finding, together with the highly
 sub-Eddington regime of accretion onto the central massive black
 hole\footnote{The $Chandra$ measurement of the nuclear 2--10 keV
 luminosity (Sect. 3.1.2) together with the black hole mass estimate
 of Kormendy et al. (1996) give $L_X\sim 10^{-7} L_{Edd}$.}, has
 encouraged the developement of advection dominated accretion flow
 (ADAF) models for LLAGNs (Quataert, Di Matteo \& Narayan 1999).  Recent
 modifications of the standard ADAF model including outflows or
 convection (Narayan, Igumenischev \& Abramowicz 2000) allow the
 luminosity to be further reduced and have been recently advocated to
 explain the very low $Chandra$ upper limits on the nuclear emission
 in three giant elliptical galaxies (Loewenstein et al. 2001). 

\subsection{NGC 4736}

From our analysis it turns out that:

\par\noindent 
1) the LECS and MECS spectra require a hard component (a power law of
$\Gamma \sim 1.6$), a soft thermal component ($kT=0.2-0.6$ keV) 
contributing $\lsim 30$\% of the
total observed flux below 2 keV, and absorption consistent with the
Galactic value.  The ACIS-S spectrum of the central $\sim 1^{\prime}$
radius region requires a two temperature fit for the soft thermal
component.

\par\noindent 
2) no Fe-K emission line is found superimposed on the hard continuum.
No short or long term flux variability is detected.

\par\noindent 3) the ACIS-S image reveals a cluster of at least four
bright point sources within the central circle of $5^{\prime\prime}$
radius, embedded in diffuse emission. The brightest source has L(2--10
keV)$\sim 3\times 10^{39}$\es\sp and $\Gamma\sim 1.2$. So, it is
similar to the most luminous [L(2--10 keV)$\gsim 10^{39}$\es] sources
discovered with ACIS-S in the starforming regions of the Antennae
galaxies (Fabbiano, Zezas \& Murray 2001; Zezas et al. 2001).  X-ray sources
with luminosities equivalent to the entire Eddington luminosity for
accretion onto a $(10-10^3) M_{\odot}$ object, recently
collectively defined
`ultra-luminous X-ray sources' (e.g., Makishima et al. 2000),
are preferentially found in starforming regions. For them, mild X-ray
beaming during a short but common stage in the evolution of
intermediate and high mass XRB's has been recently suggested (King et
al. 2001).
 
\par\noindent 
4) the compact radio nuclear
source does not coincide with the brightest X-ray source but with a fainter
[L(2--10 keV)$\sim 6\times 10^{38}$\es]
and absorbed source. Its spectrum further requires a thermal
($kT=0.6-0.7$ keV) component, perhaps a local
enhancement of hot gas density at the position of this nucleus.
The same requirement has been found studying the spectra of
 a few point sources in the star-forming regions of the Antennae
(Fabbiano et al. 2001, Zezas et al. 2001) and of the nuclear source
in the Sa galaxy NGC 1291 (Irwin, Sarazin \& Bregman 2001).

\par\noindent
5) the 1--7 keV brightness profile looks more clumpy than that of the
Sombrero galaxy (due to a larger number of bright XRBs in the central region)
 and the nuclear point source is much less dominating.

From the present investigation a starburst-induced origin for the
LINER activity is clearly preferred.  This is based on the X-ray
results showing that: 1) the presence of a LLAGN is not conspicuous; 2) there
is a large number (at least with respect to the case of Sombrero) of
hard and bright point sources at the galactic center, at least two of which
have properties similar to those of the Antennae sources (Fabbiano et
al. 2001); 3) the total 2--10 keV MECS emission 
($\sim 8\times 10^{39}$ \es) is clearly in excess of that expected for
hard binaries in normal spirals (this, calculated as in Sect. 5.1, is
$3.4\times 10^{39}$\es\sp from the Fabbiano et al. relation and
$1.1\times 10^{39}$\es\sp from the Roberts \& Warwick relation), and
this is not due to the presence of a bright nucleus, as the ACIS-S
image reveals; 4) for the soft thermal emission both $BeppoSAX$
 and ACIS-S indicate a temperature larger than the gas virial temperature
[$\sigma=136$ km s$^{-1}$ for this galaxy (McElroy 1995), that
corresponds to $kT\sim 0.1$ keV], a property that 
can be easily explained with heating of the interstellar gas
by the starburst (e.g., Heckman 2000).

Is there some residual evidence in favour of a LLAGN?  The brightness
temperature of the nuclear radio source suggests the presence of an
AGN rather than, e.g., an HII region (Turner \& Ho 1994), while the
nuclear bright UV point source revealed by $HST$ could even be a
compact star cluster (Maoz et al. 1995).  Unless the X-ray/radio
source was in a very faint state during the ACIS-S pointing [at 90\%
confidence its L(2--10 keV)$<10^{39}$\es] this possible AGN is
of extremely low luminosity.  A central supermassive black hole of
$2.2\times 10^7M_{\odot}$ is inferred from the $M_{BH}$--$\sigma$
relation (Ferrarrese \& Merritt 2000), from which we derive L(2--10
keV)/$L_{Edd}\sim 2\times 10^{-7}$.  Garcia et al. (2000) report a
similarly low luminosity X-ray source [L(0.3--7 keV)$<1.6\times
10^{38}$\es] detected by ACIS-I within $1^{\prime\prime}$ of the radio
nucleus of the M31 galaxy, which hosts a central compact dark object
of $3.0\times 10^7M_{\odot}$. This source seems to be much softer
(its $\Gamma=4.5\pm 1.5$) than the nuclear X-ray/radio source in
NGC 4736, though.

\section{Conclusions}

We have presented the analysis of the $BeppoSAX$ observations of two
nearby LINER galaxies, complemented with a study of their central
regions based on $Chandra$ ACIS-S data.  This analysis has shown that
1) the X-ray emission has a different origin in the two galaxies: an
AGN in Sombrero and a recent starburst in NGC 4736, which reinforces the
evidence that LINERs may represent more than one physical phenomenon,
as mentioned in Sect. 1;  and 2) hard
X-ray data taken at very low angular resolution (such as those that
can be obtained by $BeppoSAX$ or $ASCA$) can be misleading, and the
high resolution imaging spectroscopy provided by $Chandra$ is
essential to draw conclusions concerning the origin of the LINER
activity.  Also, the global galactic spectral properties can be very
different from those of the nuclear sources, because these do not
largely dominate the total emission.

The $BeppoSAX$ data indicate the presence of hard and extended
emission, in excess of that expected for the collective emission of
discrete X-ray sources in spirals, plus soft thermal emission, that is
more evident in the spectrum of NGC 4736.  The soft emission could
plausibly come from hot gas retained by the galactic potential well in
Sombrero, and by gas heated by supernovae in NGC 4736. An extremely
low luminosity AGN could still be present in NGC 4736, though, due to
the presence of a compact non-thermal radio source coincident with an
X-ray faint central point source.

\begin{acknowledgements}
G. Fabbiano acknowledges support from NASA contract NAS 8-39073 
(Chandra X-ray Center) and S. Pellegrini and G. Trinchieri from ASI and MURST.

\end{acknowledgements}

\end{document}